\documentclass{vow2008}

\title{Using Virtual Observatory techniques to search for Adaptive Optics suitable AGN}
\author[1]{Jens Zuther}
\author[1,2]{Gerard Lemson}
\author[3,4]{Andreas Eckart}
\author[1,5]{Wolfgang Voges}
\author[6]{Dimitri A. Gadotti}
\author[1]{Jai Won Kim}

\affil[1]{Max-Planck Institut f\"ur extraterrestrische Physik, Garching bei M\"unchen, Germany}
\affil[2]{Astronomisches Rechen-Institut, Heidelberg, Germany}
\affil[3]{I. Physikalisches Institut, Universit\"at zu K\"oln, K\"oln, Germany}
\affil[4]{Max-Planck Institut f\"ur Radioastronomie, Bonn, Germany}
\affil[5]{Max-Planck Digital Library, M\"unchen, Germany}
\affil[6]{Max-Planck Institut f\"ur Astrophysik, Garching bei M\"unchen, Germany}

\usepackage{graphicx}
\usepackage{natbib}
\begin{document}

\keywords{AGN; X-rays; Adaptive Optics; Virtual Observatory}

\maketitle

\begin{abstract}
Until recently, it has been possible only for nearby galaxies to study the scaling relations between central black hole and host galaxy in detail. Because of the small number densities at low redshift, (luminous) AGN are underrepresented in such detailed studies. The advent of adaptive optics (AO) at large telescopes helps overcoming this hurdle, allowing to reach small linear scales over a wide range in redshift. Finding AO-suitable targets, i.e., AGN having a nearby reference star, and carrying out an initial multiwavelength classification is an excellent use case for the Virtual Observatory. We present our Virtual-Observatory approach to select an AO-suitable catalog of X-ray-emitting AGN at redshifts $0.1<z<1$.
\end{abstract}

\section{Introduction}
Extensive observational work on local galaxies (passive and active) has revealed significant correlations between the mass of the central black hole and properties of its host galaxy \citep[e.g., bulge luminosity, bulge stellar velocity dispersion;][]{2003ApJ...589L..21M}. Corresponding theoretical work tries to explain the observations in an evolutionary context. The detailed physics, however, can only be studied in the nearest galaxies, since atmospheric turbulences limit the resolving power of large telescopes and, therefore, achievement of small linear scales. From space, the Hubble Space Telescope has successfully extended this work, however, on the cost of sensitivity because of the small light-collecting area. Small linear scales are necessary to properly sample nuclear and host properties. At correspondingly small cosmological distances, the number densities of AGN (and their various subclasses) is low and larger cosmological volumes have to be probed to approach meaningful statistics. Because of surface-brightness dimming and increasing linear scales, large aperture telescopes are required that can overcome the limitations imposed by the atmosphere. This can be achieved with Adaptive Optics \citep[AO; ][]{1993ARA&A..31...13B}.

AO requires a bright point source reference (guide star, GS) close to the science target. Such a GS can be a real star (natural GS, NGS) or a light spot produced by a laser beacon in the upper atmosphere (laser GS, LGS). All major large telescopes today are equipped with AO systems that can handle NGS and LGS, though, the latter systems have become available only recently. For typical constraints (see below), the sky coverage in NGS mode is only about a few percent \citep{1999aoa..book.....R}. Therefore, much of 'traditional' astronomical research is not concerned with AO. With LGS systems at hand, the setback of a small sky coverage can be overcome, however, another point source reference for tip-tilt correction is required \citep{1993ARA&A..31...13B}.

The Virtual Observatory (VObs), especially the homogeneous access to large area sky surveys, like the Sloan Digital Sky Survey \citep[SDSS;][]{2000AJ....120.1579Y}\footnote{ http://www.sdss.org}, ROSAT All-Sky Survey \citep[RASS;][]{1999A&A...349..389V}\footnote{ http://www.xray.mpe.mpg.de/cgi-bin/rosat/rosat-survey}, and cross-matching capabilities (e.g. OpenSkyQuery\footnote{http://www.openskyquery.net}), allow to search for relevant sets of AO suitable targets. 

\section{Why near-infrared?}
\label{sec:NIR}
Current AO systems work \citep[e.g.,][]{2005sao..conf.....B, 2005AJ....130.1472P, 2008MNRAS.385.1129R} in the near-infrared (NIR). This is related to technological limitations, since the number of correcting elements and the control frequency both are related to the observing wavelengths \citep[$\sim\lambda^{6/5}$;][]{1999aoa..book.....R}. The quality of the correction is a function of angular separation of the science target (usually on-axis) and the NGS, as well as, the brightness of the NGS. For the sake of a larger sample, we applied rather weak constraints on the NGSs:
\begin{itemize}
\item Science target / NGS angular separation $\leq 40\arcsec$ and 
\item NGS visual magnitude $V\leq 14$.
\end{itemize}

\begin{figure}
\centering
%\vspace{4cm}
\includegraphics[width=6cm]{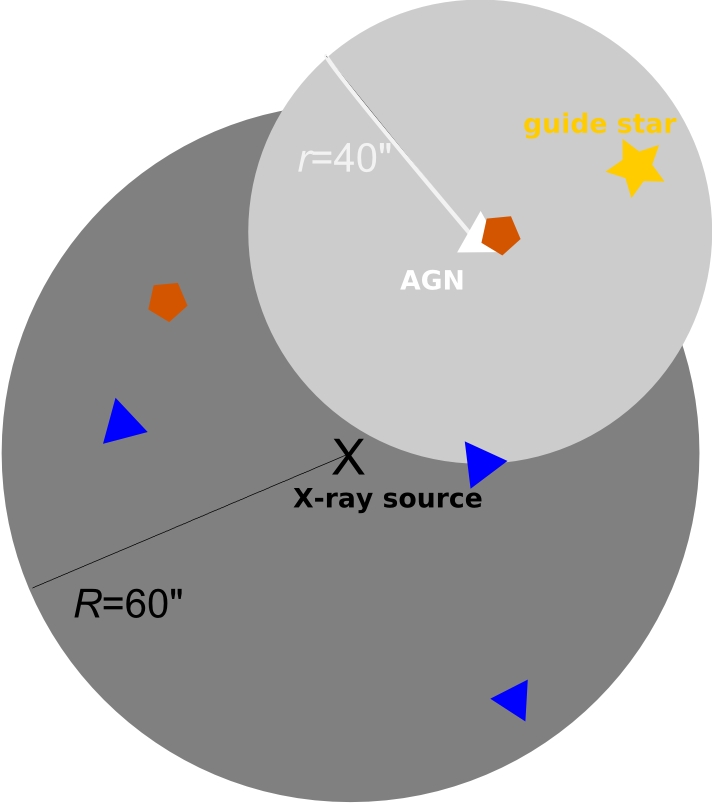}%matchingProblem.eps}
\caption{Schematic representation of the matching problem \citep[adapted from ][]{zutherThesis}.\label{fig:matchingProblem}}
\end{figure}

Further reasons of going to the NIR are of a more astrophysical nature: (i) The NIR is much less prone to dust extinction than the visible, and (ii) the contrast between the bright AGN and the host galaxy is small, as the AGN spectral energy distribution (SED) has a local minimum at around 1\,$\mu$m, whereas the host SED has a local maximum in the NIR \citep[e.g.,][]{1997quho.conf...45M}. These characteristics allow for an easier separation of AGN and host emission and a deeper probing of the circumnuclear environment.

\section{The matching problem}
In order to find a statistically meaningful number of AO-suitable targets and because of homogeneity, we decided to use large-area (if not all-sky) surveys. The combination of X-ray and visible databases has proven to allow for such a search in an efficient manner \citep{2005ARA&A..43..827B}. The RASS and the SDSS appear to be well matched in terms of sensitivity and typical X-ray-to-visible flux ratios \citep{2007AJ....133..313A}. The central problem in the SDSS/RASS matching process is the large positional uncertainty of the RASS. The average RASS error cone has a radius of 1\arcmin\, \citep[e.g.,][]{2008AJ....135...10P}. This leads to the combination of catalogs of different resolution. The situation is depicted in Fig. \ref{fig:matchingProblem}. On average we find about 7 SDSS objects (with $r<19$) within 1\arcmin\, of the X-ray position.
\begin{figure*}
\centering
%\vspace{4cm}
\includegraphics[width=13cm]{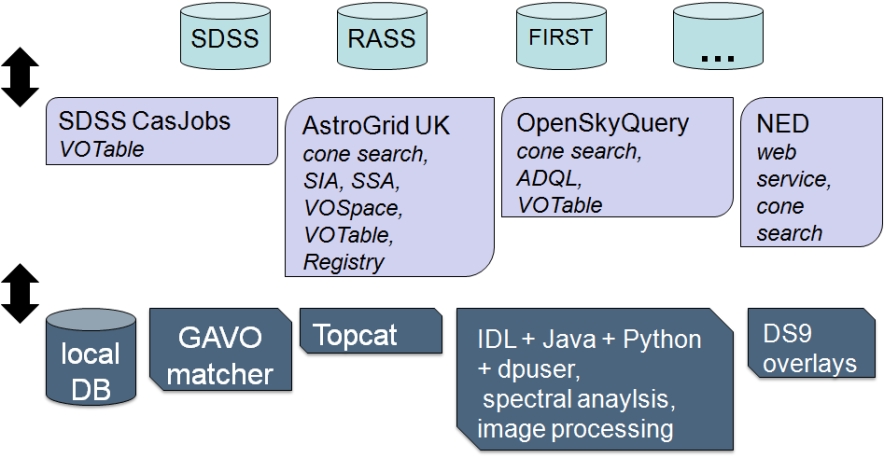}%vo_tools.eps}
\caption{Three layers comprising the VObs. The top level consists of the original datasets, like SDSS, RASS, etc. The 'core' layer is some kind of middleware that allows the clients/users - at the bottom level - to access the astronomical archives in a standardized way.\label{fig:voTools}}
\end{figure*}
Therefore, it is not easy to decide, which is the 'true' visible counterpart of the X-ray source \citep[cf. discussion in][]{2008AJ....135...10P}. 

Here we focus on the VObs tools that can help in the matching process. The details will be the topic of a forthcoming article. Figure \ref{fig:voTools} presents the publicly available and custom made tools. Many of these are aware of middleware based on standards provided by the IVOA\footnote{International Virtual Observatory Alliance, http://www.ivoa.net}, such as simple image access (SIA), simple cone search (SCS), or simple spectra access (SSA). Cross-matching primarily involves SCS. Interoperability between the individual tools is another important aspect of the VObs. Currently, the direct transfer of data in the VOTable format via PLASTIC\footnote{Platform for Astronomy Tool InterConnection: http://www.ivoa.net/Documents/latest/PlasticDesktopInterop.html} is the primary means of such interoperability. However, we set the VObs scheme in a broader context, including tools in a programmable environment, which are not necessarily VObs compatible in the IVOA sense. Such an environment (e.g., Python, Java, IDL) allows for the required interoperability across all the tools.

\begin{figure}[h!]
\centering
%\vspace{4cm}
\includegraphics[width=8cm]{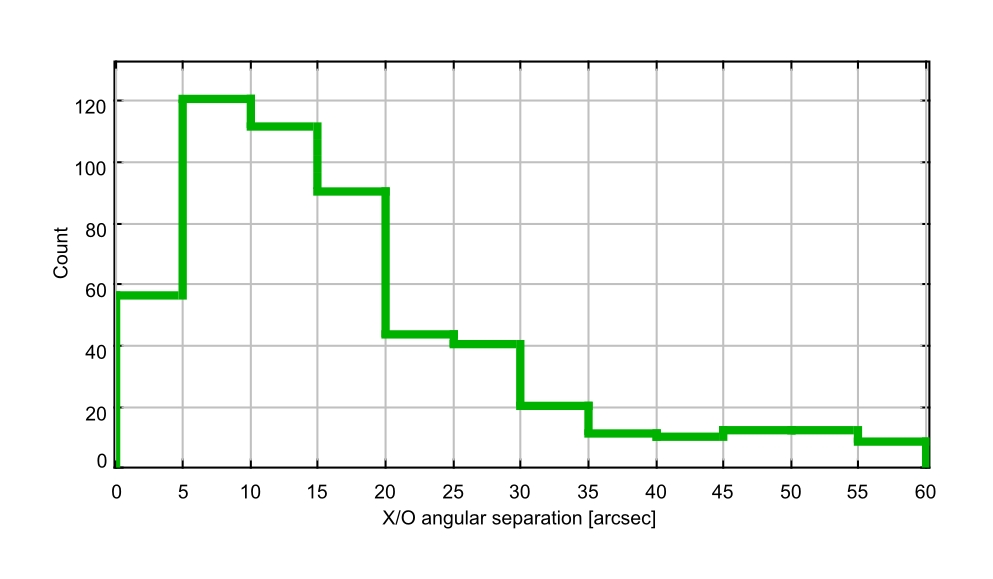}%xo_angular_separation.eps}
\caption{Histogram of X-ray/visible angular separations in arcsec.\label{fig:angSep}}
\end{figure}

Since we are interested in AGN and pairs of AGN/X-ray sources that have a high matching probability, we use the pragmatic approach of selecting those visible candidates that have an SDSS spectrum (classified either as galaxy or AGN). Typically, only one spectroscopic SDSS object is found within the RASS error cone. Of the order of a few percent of these candidates have a nearby NGS according to the criteria in the Sec. \ref{sec:NIR}. The X-ray/visible angular separation histogram of the remaining candidates (Fig. \ref{fig:angSep}) displays the typical behavior compared to other work \citep{2007AJ....133..313A,2008AJ....135...10P}, i.e. a strong correlation of the X-ray and visible positions with a peak at $\sim 10\arcsec$. \citeauthor{2008AJ....135...10P} find from their Monte Carlo simulations that the matching fraction for AGN is $>80$\% for angular separations $<30$\arcsec. More details on probabilistic cross-matching can be found in these proceedings.

In addition to these criteria we cross-checked our results by cross-matching the data set with NED\footnote{NED: http://nedwww.ipac.caltech.edu/ via the 
NED web service at http://voservices.net/NED/ws\_v2\_0/NED.asmx} and catalogs at different wavelengths like 20\,cm radio surveys (FIRST, NVSS) and NIR (2MASS). For the latter catalogs we used our Java-based cross-matching tool \citep[cf.][]{2006ASPC..351..695A}, which accesses the Vizier database\footnote{http://webviz.u-strasbg.fr/viz-bin/VizieR}. NED provides a comprehensive, but not homogeneous, source description based on references in the literature. One interesting case in our data set is 3C~273, in which the QSO has no SDSS spectrum, but was chosen to be the NGS - as it is unresolved in the SDSS - of a less luminous AGN in the ROSAT cone. Such situation can only be resolved by using an VObs approach as described above, in order to achieve as much information as possible.

\section{Outlook}
\begin{figure}[h!]
\centering
%\vspace{4cm}
\includegraphics[width=8cm]{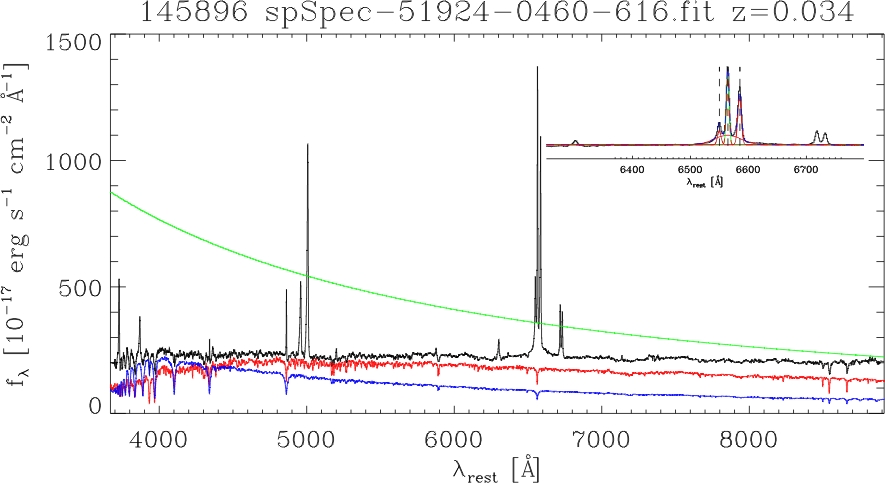}%mrk609_sdss_spec.eps}
\caption{SDSS spectrum of Mrk~ 609. Shown are the individual components that are used for the fitting of the continuum (green: power-law, red: young starburst, blue: old starburst) and the emission lines (inset).\label{fig:mrk609spec}}
\end{figure}

Based on the matched sample, we have carried out a rough spectral analysis of the candidates \citep[cf.][]{zutherThesis}. We fitted the spectra with a linear combination of a power-law, a young and an older stellar population, and allowing for dust extinction (Fig. \ref{fig:mrk609spec}). In the 'continuum subtracted' spectra, we then fitted narrow and broad components of prominent emission lines (Fig. \ref{fig:mrk609spec} inset). We used custom-made IDL code for this, but used available data access methods to the SDSS archive and our local, relational sample database. The results, which include power-law index, reddening, fractions of AGN, young and old starburst, emission line width and fluxes, Balmer ratios, etc., are also stored in the database.
In a future step, we aim at making this database VObs compliant (i.e., primarily include metadata) and then publicly accessible.

\begin{figure*}
\centering
%\vspace{4cm}
\includegraphics[width=13cm]{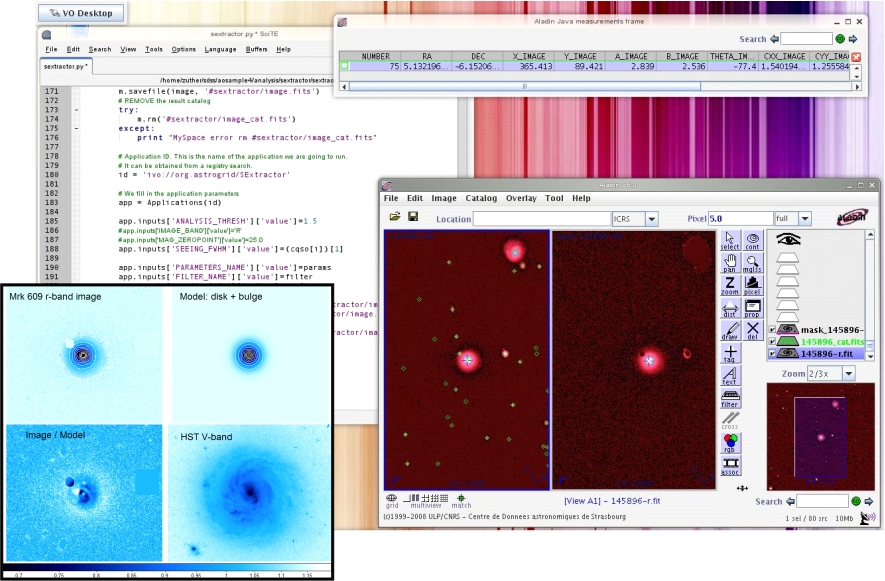}%astrogrid_budda.eps}
\caption{Prototype of BUDDA in the VObs framework. Python scripting within the AstroGrid VODesktop allows to call the SExtractor web service, to display results in Aladin, and to pipe the results into BUDDA. The 2D decomposition is shown in the left inset for Mrk~609 (SDSS input, bulge + disk model, residual, HST V-band image).\label{fig:budda}}
\end{figure*}

Besides the matching and photometric/spectroscopic analysis, we are currently working on a VObs compliant workflow concerning the morphological decomposition of imaging data using BUDDA \citep{2008MNRAS.384..420G}. 2D decomposition allows for a better disentangling of nuclear and host emission components. This is important when, e.g., searching for substructures, which might be related to AGN fueling, like inner bars, inner spirals, etc. Figure \ref{fig:budda} presents a prototype that makes use of the AstroGrid VODesktop\footnote{http://www.astrogrid.org}. VODesktop unifies the discovery of data and services via registries with an online storage (VOSpace), a facility to create workflows, and provides scripting capabilities. As a preparatory step, images to be fed to BUDDA have to be cleaned of extra-target sources. This task can be accomplished with SExtractor \citep{1996A&AS..117..393B} in the form of a web service.

Having established such a workflow, it will be possible to do a morphological decomposition of the SDSS based parent sample and also of the higher angular resolution NIR images of the follow-up programs. 

Tools that allow batch processing within the VObs framework will become essential for the next generation of telescopes like the Large Binocular Telescope (LBT). It will make use of multi-conjugate AO, requiring several GSs within the field-of-view \citep[cf.][]{2008SPIE.7014E..43H}, also requiring more advanced cross-matching schemes. The LBT will produce large and complex data sets \citep[cf.][]{2004SPIE.5491..106E} that can only be analyzed in highly automatized workflows, which will encompass information and services from the VObs.

\section*{Acknowledgments}
Jens Zuther acknowledges support from the European project EuroVO-DCA under the Research e-Infrastructures area.

\bibliographystyle{aa}
\bibliography{zuther}

\begin{thebibliography}{18}
\expandafter\ifx\csname natexlab\endcsname\relax\def\natexlab#1{#1}\fi

\bibitem[{{Adorf} {et~al.}(2006){Adorf}, {Lemson}, \&
  {Voges}}]{2006ASPC..351..695A}
{Adorf}, H.-M., {Lemson}, G., \& {Voges}, W. 2006, in ASP Conference Series,
  Vol. 351, ADASS XV, ed. C.~{Gabriel}, C.~{Arviset}, D.~{Ponz}, \&
  S.~{Enrique}, 695

\bibitem[{{Anderson} {et~al.}(2007){Anderson}, {Margon}, {Voges}, {Plotkin},
  {Syphers}, {Haggard}, {Collinge}, {Meyer}, {Strauss}, {Ag{\"u}eros}, {Hall},
  {Homer}, {Ivezi{\'c}}, {Richards}, {Richmond}, {Schneider}, {Stinson},
  {Vanden Berk}, \& {York}}]{2007AJ....133..313A}
{Anderson}, S.~F., {Margon}, B., {Voges}, W., {et~al.} 2007, \aj, 133, 313

\bibitem[{{Beckers}(1993)}]{1993ARA&A..31...13B}
{Beckers}, J.~M. 1993, \araa, 31, 13

\bibitem[{{Bertin} \& {Arnouts}(1996)}]{1996A&AS..117..393B}
{Bertin}, E. \& {Arnouts}, S. 1996, \aaps, 117, 393

\bibitem[{{Brandner} \& {Kasper}(2005)}]{2005sao..conf.....B}
{Brandner}, W. \& {Kasper}, M.~E., eds. 2005, {Science with Adaptive Optics}

\bibitem[{{Brandt} \& {Hasinger}(2005)}]{2005ARA&A..43..827B}
{Brandt}, W.~N. \& {Hasinger}, G. 2005, \araa, 43, 827

\bibitem[{{Eckart} {et~al.}(2004){Eckart}, {Zuther}, {Mouawad}, {Schodel},
  {Straubmeier}, {Bertram}, {Pott}, {Scharwachter}, \&
  {Herbst}}]{2004SPIE.5491..106E}
{Eckart}, A., {Zuther}, J., {Mouawad}, N., {et~al.} 2004, in SPIE Conference
  Series, ed. W.~A. {Traub}, Vol. 5491, 106

\bibitem[{{Gadotti}(2008)}]{2008MNRAS.384..420G}
{Gadotti}, D.~A. 2008, \mnras, 384, 420

\bibitem[{{Herbst} {et~al.}(2008){Herbst}, {Ragazzoni}, {Eckart}, \&
  {Weigelt}}]{2008SPIE.7014E..43H}
{Herbst}, T.~M., {Ragazzoni}, R., {Eckart}, A., \& {Weigelt}, G. 2008, in SPIE
  Conference Series, Vol. 7014

\bibitem[{{Marconi} \& {Hunt}(2003)}]{2003ApJ...589L..21M}
{Marconi}, A. \& {Hunt}, L.~K. 2003, \apjl, 589, L21

\bibitem[{{McLeod}(1997)}]{1997quho.conf...45M}
{McLeod}, K.~K. 1997, in Quasar Hosts, ed. D.~L. {Clements}, 45

\bibitem[{{Parejko} {et~al.}(2008){Parejko}, {Constantin}, {Vogeley}, \&
  {Hoyle}}]{2008AJ....135...10P}
{Parejko}, J.~K., {Constantin}, A., {Vogeley}, M.~S., \& {Hoyle}, F. 2008, \aj,
  135, 10

\bibitem[{{Prieto} {et~al.}(2005){Prieto}, {Maciejewski}, \&
  {Reunanen}}]{2005AJ....130.1472P}
{Prieto}, M.~A., {Maciejewski}, W., \& {Reunanen}, J. 2005, \aj, 130, 1472

\bibitem[{{Riffel} {et~al.}(2008){Riffel}, {Storchi-Bergmann}, {Winge},
  {McGregor}, {Beck}, \& {Schmitt}}]{2008MNRAS.385.1129R}
{Riffel}, R.~A., {Storchi-Bergmann}, T., {Winge}, C., {et~al.} 2008, \mnras,
  385, 1129

\bibitem[{{Roddier}(1999)}]{1999aoa..book.....R}
{Roddier}, F. 1999, {Adaptive optics in astronomy} (Cambridge University Press)

\bibitem[{{Voges} {et~al.}(1999){Voges}, {Aschenbach}, {Boller},
  {Br{\"a}uninger}, {Briel}, {Burkert}, {Dennerl}, {Englhauser}, {Gruber},
  {Haberl}, {Hartner}, {Hasinger}, {K{\"u}rster}, {Pfeffermann}, {Pietsch},
  {Predehl}, {Rosso}, {Schmitt}, {Tr{\"u}mper}, \&
  {Zimmermann}}]{1999A&A...349..389V}
{Voges}, W., {Aschenbach}, B., {Boller}, T., {et~al.} 1999, \aap, 349, 389

\bibitem[{{York} {et~al.}(2000){York}, {Adelman}, {Anderson}, {Anderson},
  {Annis}, {Bahcall}, {Bakken}, {Barkhouser}, {Bastian}, {Berman}, {Boroski},
  {Bracker}, {Briegel}, {Briggs}, {Brinkmann}, {Brunner}, {Burles}, {Carey},
  {Carr}, {Castander}, {Chen}, {Colestock}, {Connolly}, {Crocker}, {Csabai},
  {Czarapata}, {Davis}, {Doi}, {Dombeck}, {Eisenstein}, {Ellman}, {Elms},
  {Evans}, {Fan}, {Federwitz}, {Fiscelli}, {Friedman}, {Frieman}, {Fukugita},
  {Gillespie}, {Gunn}, {Gurbani}, {de Haas}, {Haldeman}, {Harris}, {Hayes},
  {Heckman}, {Hennessy}, {Hindsley}, {Holm}, {Holmgren}, {Huang}, {Hull},
  {Husby}, {Ichikawa}, {Ichikawa}, {Ivezi{\'c}}, {Kent}, {Kim}, {Kinney},
  {Klaene}, {Kleinman}, {Kleinman}, {Knapp}, {Korienek}, {Kron}, {Kunszt},
  {Lamb}, {Lee}, {Leger}, {Limmongkol}, {Lindenmeyer}, {Long}, {Loomis},
  {Loveday}, {Lucinio}, {Lupton}, {MacKinnon}, {Mannery}, {Mantsch}, {Margon},
  {McGehee}, {McKay}, {Meiksin}, {Merelli}, {Monet}, {Munn}, {Narayanan},
  {Nash}, {Neilsen}, {Neswold}, {Newberg}, {Nichol}, {Nicinski}, {Nonino},
  {Okada}, {Okamura}, {Ostriker}, {Owen}, {Pauls}, {Peoples}, {Peterson},
  {Petravick}, {Pier}, {Pope}, {Pordes}, {Prosapio}, {Rechenmacher}, {Quinn},
  {Richards}, {Richmond}, {Rivetta}, {Rockosi}, {Ruthmansdorfer}, {Sandford},
  {Schlegel}, {Schneider}, {Sekiguchi}, {Sergey}, {Shimasaku}, {Siegmund},
  {Smee}, {Smith}, {Snedden}, {Stone}, {Stoughton}, {Strauss}, {Stubbs},
  {SubbaRao}, {Szalay}, {Szapudi}, {Szokoly}, {Thakar}, {Tremonti}, {Tucker},
  {Uomoto}, {Vanden Berk}, {Vogeley}, {Waddell}, {Wang}, {Watanabe},
  {Weinberg}, {Yanny}, \& {Yasuda}}]{2000AJ....120.1579Y}
{York}, D.~G., {Adelman}, J., {Anderson}, Jr., J.~E., {et~al.} 2000, \aj, 120,
  1579

\bibitem[{{Zuther}(2007)}]{zutherThesis}
{Zuther}, J. 2007, PhD thesis, University of Cologne, Cologne, Germany

\end{thebibliography}

\end{document}